\documentclass{article}
\usepackage{amsmath,amssymb,graphicx,mathrsfs,hyperref,slashed}
\usepackage[titletoc,toc,title]{appendix}
\usepackage{xcolor}
\newcommand{\be}{\begin{equation}}
\newcommand{\ee}{\end{equation}}
\newcommand{\bea}{\begin{eqnarray}}
\newcommand{\eea}{\end{eqnarray}}

\def\({\left(} \def\){\right)}

\renewcommand{\baselinestretch}{1.5}
\begin{document}
\title{\vspace{-1.8in}
{Black holes and information: A new take on an old paradox}}
\author{\large K.L.H. Bryan${}^{(1,2)}$,  A.J.M. Medved${}^{(1,2)}$
\\
\vspace{-.5in} \hspace{-1.5in} \vbox{
 \begin{flushleft}
$^{\textrm{\normalsize (1)\ Department of Physics \& Electronics, Rhodes University,
  Grahamstown 6140, South Africa}}$
$^{\textrm{\normalsize (2)\ National Institute for Theoretical Physics (NITheP), Western Cape 7602,
South Africa}}$
\\ \small \hspace{1.07in}
  g08b1231@gmail.com,\  j.medved@ru.ac.za
\end{flushleft}
}}
\date{}
\maketitle

\begin{abstract}

Interest in  the  black hole information paradox  has recently  been catalyzed by the  newer ``firewall'' argument. 
The crux of the updated argument is that previous solutions which relied  on  observer complementarity are in violation of  
the quantum condition of   monogamy of entanglement; with the prescribed remedy being to discard the equivalence principle 
in favor of an energy barrier (or firewall) at the black hole horizon. Differing points of view have been  put forward, 
including  the ``ER=EPR'' counterargument and  the final-state solution, both of which can be viewed as  potential 
resolutions to the apparent conflict between quantum monogamy and Einstein equivalence. After reviewing these recent 
developments, this paper argues that the  ER=EPR   and  final-state solutions can --- thanks to observer complementarity --- 
be seen as the same resolution of the  paradox but from two different  perspectives: inside and outside the black hole.

\end{abstract}
\newpage
\renewcommand{\baselinestretch}{1.5}\normalsize

\section{Introduction}

  Black holes have provided an alluring yet confusing arena for the study of physics. One suddenly encounters paradoxes when  
  standard concepts, which  are taken for granted in
  other  physical environments, are applied  to a black hole and its surroundings.
  A particularly  notorious paradox is the apparent  destruction of information when matter
  falls through a  black hole horizon~\cite{Hawk_info}. This loss of information presents a direct challenge to the principles 
  of quantum theory and, although it has been the subject of  intense scrutiny, 
  the puzzle continues to persist.

There once was a commonly held viewpoint that the  information
paradox could be resolved  by virtue of a framework that is  known
--- in the spirit of Bohr ---  as horizon or observer  complementarity~\cite{suss}.
However,  a recent addition to this debate  suggests that the  information
in question never
does  makes it past the horizon.  Rather, the black hole  horizon  is
surrounded  by  a high-energy barrier
--- or   ``firewall'' ---  which thermalizes in-falling  matter on
impact~\cite{amps}.
The firewall  argument
  is based on an apparent violation   of the quantum condition
of  ``monogamy of entanglement''  but is by no means generally accepted.
This is because the loss of quantum monogamy
would be no less costly  than giving up  the long-cherished  equivalence principle
of Einstein relativity. The latter principle dictates
that one encounters  an approximately flat spacetime when falling through the 
horizon.

  One viable resolution of  the firewall puzzle is the 
``ER=EPR'' counterargument as put forth by  Maldacena and Susskind~\cite{maldacenaSuss}. Those authors 
adopt the same basic argument as the firewall proponents but draw a different conclusion. In order to preserve
  a natural state of approximately flat space at the horizon,
Maldacena and Susskind suggest that  wormholes ({\em i.e.}, Einstein--Rosen or ER bridges)
  allow for disturbances to travel between the  Hawking-radiated matter \cite{hawk74}   and   the environment
within the black hole interior.

Another possible resolution is the so-called final-state solution,
which was  first put forward as a solution to the information paradox by
  Horowitz and Maldacena~\cite{horow}.
but latter updated for the firewall scenario by Lloyd and Preskill~\cite{lloydAMPS}.  The proposed  procedure allows   
information  to escape the
  black hole through the post-selection  of a  specific final state
at the black hole  singularity. In effect, a process of  quantum teleportation
  in used  to transfer the state of the in-falling matter to that of
the exterior
radiation. But, as post-state teleportation is formally no different than
the propagation of quantum information through wormholes~\cite{lloydPST},
the two discussed solutions  --- ER=EPR and final state --- would appear
to share some similarities on at least a superficial level. Here, we will make a much stronger
claim.

  The conclusion in this paper is
that the  ER=EPR and final-state solutions are simply two sides of the same
argument.  Each is reached separately on the basis of  the position
  of the observer whose perspective is in question. In both cases, a wormhole serves as the conduit for information  
  transfer but, in one case, the information travels out of the black hole and, in the other, the information
  is rather transmitted inward.
The key to this identification lies
within the auspices of  observer complementarity, which then
remains central to the information paradox in spite of its recent
detractors.

Although wormholes  are suggestive of their  own special brand of
grand-paternal-like paradoxes and associated  violations
of causality, these conflicts  are, as discussed later,  only apparent and   
resolved within the relevant theories.
 
\section{Black Holes and Information}

  Black holes were originally thought to consume all matter which fell past their horizons, without
  any hope for recovery. Essentially, they acted as cosmic
  drains in spacetime, permanently removing  matter and energy from the Universe.
This point of view was, however,   challenged 
 by Hawking, who  showed that black holes evaporate over time and  eventually disappear after being converted to  
 radiation~\cite{hawk74}. This  evaporation process meant that black holes could, after all,  restore matter and energy to the Universe.
The information content  that was  carried by the 
in-falling matter was, however, quite another story.

Black holes came to be be viewed, rather than drains, as scrambling machines.
Indeed, even an initially pure state of collapsing matter
would apparently be converted to a mixed state when it 
reemerged in the outgoing radiation~\cite{Hawk_info}. The problem
  of black holes then took on a different perspective: It appeared that energy
was conserved  but  information
was completely scrambled.
  This viewpoint is especially problematic when confronted with the quantum-mechanical  requirement of  unitary
evolution. A pure state that has  collapsed into a black hole must, by the tenets of quantum mechanics, be recoverable (in principle)
  from the radiation which eventually replaces the evaporating black hole.
That the information about this state (or that of any in-falling matter)
appears to be lost is what constitutes the black hole information paradox.

\begin{figure}[ht]
    \centering
      \includegraphics[width=0.4\textwidth]{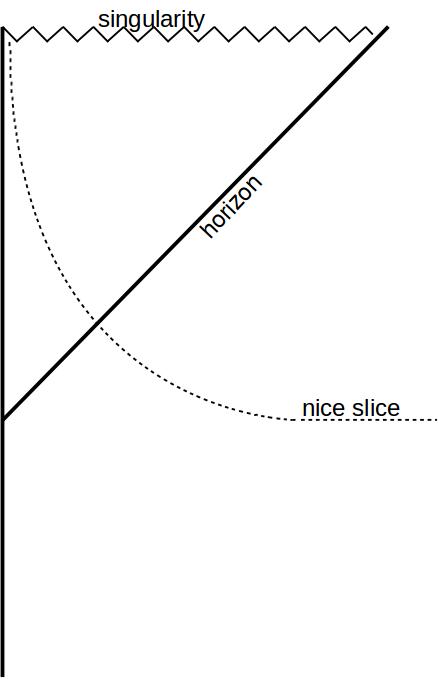}
    \caption{\em Schematic of a single ``nice slice'' crossing a black hole horizon.}
      \label{fig:niceslice}
  \end{figure}

The problem is often recast in terms of the  ``nice-slice'' point of  view (see Fig.~\ref{fig:niceslice}). 
Here,  one considers a collection of
non-intersecting and (mostly) space-like surfaces, each of which crosses the horizon such that, once inside, it  
slowly curves so as to   avoid the singularity for as long
as possible. The implication for in-falling
matter  is that there will always  be at least one nice slice which is intersected
by  both
the matter in  its original form and its reincarnation as
emitted radiation. 
If this
  radiation retains any quantum  information about the in-falling matter, then the information exists twice on the 
  same space-like slice, which directly
violates  the  so-called no-cloning principle of quantum 
mechanics~\cite{nogo}.  
For a more detailed explanation of the no-cloning theorem, please see Appendix~\ref{AppA}.
And so we are left with the paradox of either the cloning of information
  or else its destruction, with  both options being in violation
of sacred quantum principles.
 
  \subsection{Complementarity at the Horizon}
   
    A solution was offered to this problem by Susskind and collaborators,
who employed a principle that was originally  known as  horizon
    complementarity~\cite{suss}. This principle stated that two observers --- on either side of a black
    hole horizon --- may disagree on an event inasmuch as  they would never be in a position to compare their respective experiences. 
    The concept was later
expanded and renamed as observer complementarity, which follows the same principle but
 now   applying to all causally separated observers and not  just
those separated by black hole horizons~\cite{bousso}.

Regarding the black hole information paradox,  observer complementarity
allowed in-falling information to be
    cloned at the horizon without any  violation of  quantum mechanics.
The argument considers two observers: One named Alice, who falls into the black hole, and another named Bob, who remains
    outside and witnesses Alice's descent. As per Einstein's equivalence principle, 
  the details of which are expanded on in Appendix~\ref{AppB},
    Alice must experience approximately
    flat space as she enters the black hole (assuming a  large enough
black hole, as we always do). However, from  Bob's perspective, Alice  must
be  thermalized near the horizon, with her information being carried away by the  emitted radiation.  According to  horizon
    complementarity,  both events can happen. This is  because Alice and Bob can never disagree with
    one another's experience once Alice has crossed the horizon. Suppose that Bob collects Alice's radiation and then
    follows her into the black hole in an attempt to produce a paradoxical situation. Then  he can never 
 meet up with nor receive a signal from Alice before his destruction at the singularity, as such a meeting or signal  would require  
 Alice to have access to  more energy than that contained by the black hole. In this way, horizon complementarity allows cloning at 
 the horizon and, therefore,
    the preservation of any information that had passed into the black hole.
   
\section{Horizon Complementarity up in Flames}

  Although horizon complementarity was long considered a suitable resolution to the information paradox,
  the debate has been sparked anew following the ``firewall'' argument of Almheiri, Marolf, Polchinski, and Sully,
who are commonly known as
  AMPS~\cite{amps}. (Similar concerns had been expressed, but with much less fanfare, in earlier articles~\cite{itzhaki}  
  ~\cite{mathur1}  ~\cite{mathur2}  ~\cite{braunstein2009}.)

\begin{figure}[ht]
    \centering
      \includegraphics[width=0.6\textwidth]{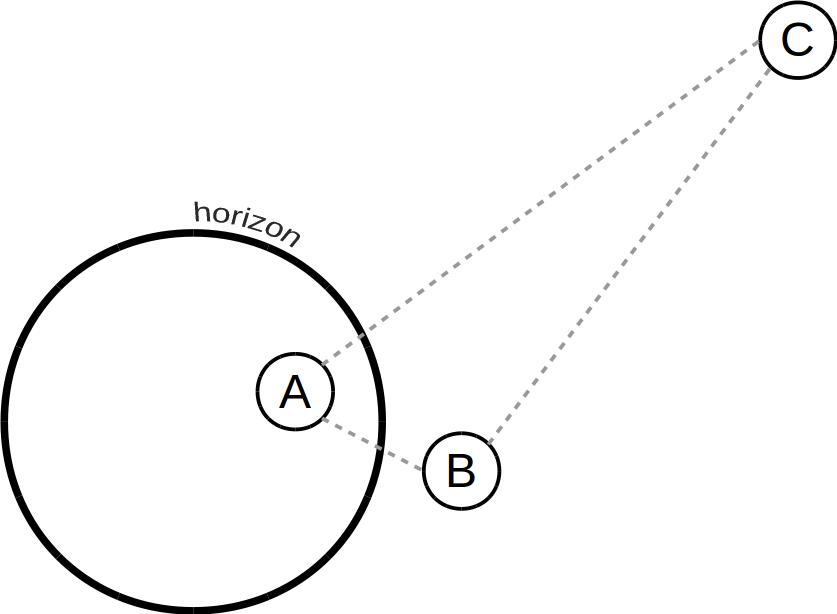}
    \caption{\em Systems involved in the firewall argument, with dotted lines showing the entanglement required by horizon complementarity}
    \label{fig:firewall}
  \end{figure}

The  firewall argument assumes a relatively old black hole~\footnote{Here, ``relatively old'' means that, for the black 
hole--radiation system, the radiation 
subsystem should be the dominant one, if only by an infinitesimal amount.\label{footnoteX}} and  begins with the acceptance 
of the same postulates as presented in the original
  horizon-complementarity framework. Those relevant to the current discussion include the expectation of
(approximately)  flat
  space for someone  crossing the horizon, any information held within the black hole can eventually be retrieved from the Hawking radiation,
  and anyone remaining  outside the black hole should see no violation of
conventional  physics.
  With these restrictions in mind, AMPS  considered three isolated matter systems, as depicted in Fig.~\ref{fig:firewall}, 
  that  are present during the evaporation
  process: One system, $A$, which lies within the black hole horizon, another system, $B$, consisting of  Hawking radiation 
  that is emitted late
 in the evaporation process   and a third system, $C$, which  is composed of  some much-earlier-emitted   Hawking radiation. 
To make the argument fully come to life, one assumes that
the collapsed matter was initially in a pure state
and that $A$ is chosen, without
loss of generality, so that its state is the purifier of $B$'s (and {\em vice versa}).

  The problem for horizon complementarity arises when the entanglement between the systems  is considered with
  reference to the aforementioned postulates. Importantly, the requirement of flat space at the horizon entails a high degree of 
  entanglement between
the interior and the modes just outside the horizon (such an entangled state for the near-horizon
region is referred to as the Unruh vacuum state~\cite{Unruh}). As a consequence,
$B$ must be highly entangled with its purifier $A$.
However, in order for an external  observer  to be able to consistently recover  information from  inside the black hole and 
--- at the same time ---  see 
nothing paradoxical, there must typically  be a high degree of entanglement
between  samples of early and late  radiation. That is,  $B$ must also be highly entangled
with $C$.~\footnote{The condition of an ``old'' black hole, as per fn.~\ref{footnoteX}, rules out the possibility that this 
conclusion can be avoided by both $B$ and $C$ being sufficiently  entangled
with systems behind the horizon.}  
This presents a situation where System~$B$ violates the principle
of  monogamy of entanglement,  which
  restricts a system to be strongly entangled with only one other system at a time (this follows directly from the strong
subadditivity of entropy~\cite{araki}
  and the argument is outlined in Appendix~\ref{AppC}). 
This violation   
would not necessarily be a problem for horizon complementarity, except
that  AMPS  confirm the existence of a  frame which allows a single  observer to witness both
  the late--early  entanglement ($B$--$C$) as well as the trans-horizon entanglement ($A$--$B$). This was the spanner in the works
  for observer complementarity, which only applies if no violation is ever
witnessed.

The resolution of the
  problem, as  prescribed by AMPS, is to 
do away with the trans-horizon entanglement and accept a sea of high-energy particles in the vicinity of the horizon. In other words,
discard  the equivalence principle in order to preserve the unitarity of the
evaporation process. However,
  it may not be necessary to throw away any sacred principles
in order  to resolve the
  black hole information paradox. All that might be needed is a wormhole.
 
\section{The ``ER=EPR'' Counterproposal}

  Although AMPS concluded that the horizon of a black hole is  a place of fiery death, Susskind and Maldacena
  proposed a different view of their argument, which might just save the equivalence principle~\cite{maldacenaSuss}.
  The crux of Susskind and Maldacena's counterargument is the
recognition of
what was   a hidden  assumption in the AMPS'  presentation.
 This assumption, to be elaborated on below,
was related to the ability of the three relevant
systems, $A$, $B$ and $C$, to transmit the effects of disturbances to one another.

Susskind and Maldacena argued that $A$, the system lying inside the
  black hole, must be in some sense ``identified'' with  $C$, the distant  system  of early Hawking radiation, in order
for both  $A$ and $C$  to be highly entangled with System~$B$ near the horizon.
However,
as Susskind and Maldacena also  point out, this identification cannot work
unless  a disturbance at $C$ directly  affects  $A$ (and {\em vice versa}).
In particular, given this identification,  a disturbance at $C$ can be expected to create particles at $A$,
which an in-falling
  observer would  view  as part of a  firewall. However, since System~$C$ was supposed to be emitted early in the evaporation process, 
  it should be too far away from the black hole
for such a disturbance to effect the journey 
  of most in-falling observers.  To this end,  AMPS  claim
  that there must always  be a  firewall  at the horizon irrespective  of any interference effects
at $C$.

  The assumption that Susskind and Maldacena took issue with is this inability for the distant radiation to rapidly transmit
  an effect to the interior of the black hole. If a disturbance of the distant radiation (or $C$) could somehow be felt by System~$A$,  then
  the effect would be to create a  firewall on the horizon for any observer in the vicinity. On the other hand, if an observer
  fell in without any interaction occurring on the distant radiation,  
  he or she  would indeed witness approximately flat space at the horizon.
But, in  neither case, would there be an observed violation of monogamy
  of entanglement  because  a distant observer
choosing to act (or not) on $C$  could never be sure about the entanglement
between $A$ and $B$, whereas
 an in-falling observer could never be sure about any relation between $B$ and $C$.

What is then  required to bypass the AMPS argument is a
  mechanism that  would allow the far-away system $C$ to transmit, almost
instantaneously, an effect to the interior system $A$. For this purpose, Susskind and Maldacena introduced the notion of Einstein--Rosen (ER) 
Bridges
  or, as they are more commonly known, wormholes. The presence of a wormhole connecting System~$C$ to System~$A$ would provide the necessary
  ``shortcut'' for any effect at $C$ to influence $A$. In this way, it can be ensured that any  firewall  would arise as the result of 
  interference on a 
near-horizon matter system (in this case, $A$) rather than  as a pre-ordained requirement of the black hole environment.
 On the other hand, such a shortcut would manifest itself as an instantaneous action at a distance,  leading to an apparent violation of 
  the principles of special relativity.  But it is, indeed,  only
an  apparent violation. A disturbance  at $C$ is transmitted
to $A$  via a legitimate
pathway though spacetime given that   a wormhole
is a direct consequence of two systems being entangled.  
This argument --- that wormholes are part and parcel
with entanglement  ---
  is known as ER=EPR,~\footnote{It should be emphasized that this
identification is conjectural and not supported by direct evidence.} where 
EPR refers to the Einstein--Podolsky--Rosen
brand  of entanglement as per the famous thought experiment~\cite{EPR}.
  Further illustration of the interpretation that ER bridges link entangled 
  systems is provided in Appendix~\ref{AppD}, where the EPR thought experiment is used  
  to demonstrate the concept further.
 This  counterargument may or may not  put the matter to rest, depending on one's taste. However, it is by no means the only workable 
 solution in the literature. Let us turn
to another.
 
\section{The Final State Solution}

 Well before firewalls and ER=EPR, there was the final-state solution, as  first proposed by Horowitz and
Maldacena~\cite{horow} and later updated by Lloyd
and Preskill~\cite{lloydAMPS}.
This proposal addressed  the black hole information paradox by employing the quantum mechanical notions of
teleportation~\cite{plenio}
and post-state selection~\cite{PSS}  as a means for transferring  information from   
  the black hole interior to particles outside the horizon. As in standard quantum teleportation, the final-state solution relies on
  an entangled pair of particles; namely, a positive- and
  negative-energy pair that initially straddles the horizon.
(In the Hawking picture of black hole
evaporation~\cite{Hawk_info},
entangled pairs are produced  at the horizon whereby the
positive-energy particle moves outward to become a quantum of Hawking radiation, while the negative-energy partner falls in and eventually 
lowers the mass
of the black hole.)

\begin{figure}[ht]
    \centering
      \includegraphics[width=0.4\textwidth]{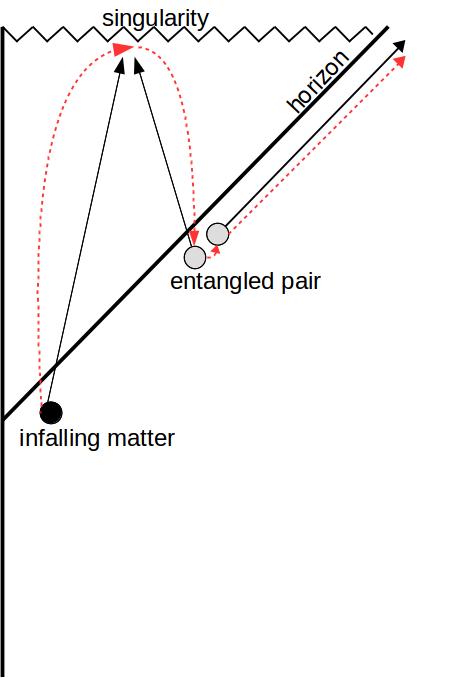}
    \caption{\em Schematic description of the final-state solution, with the red dotted line showing the path taken by 
	      the information contained in the in-fallen matter as it is teleported back in time}
    \label{fig:finalstate}
  \end{figure}

Let us, for current purposes, denote the positive-energy particle as System~2 and  its
  negative-energy partner as System~3, with  System~1
reserved for  denoting a suitable particle in the original in-falling matter system.
(This setup is illustrated in Fig.~\ref{fig:finalstate}.)
More to the point, System~1 will be the particle
that is responsible for the annihilation of System~3.
In their proposal, Horowitz and Maldacena
regard  this annihilation event as a ``measurement'' of the two particles for which a specific final-state
  is specified ({\em i.e.}, post-selected). As System~3 is entangled with System~2, the outcome of this measurement must also affect the 
  latter system.
In fact, through the combination of post-state  selection and quantum teleportation, it can be ensured that System 2, the incipient Hawking
particle,  is in the  exact same  state as that of System 1, the in-falling bit of matter.

The required protocol is essentially a special case of quantum teleportation that adopts  post-state selection as a means for  negating 
the need for any classical communication. The information is still transferred
  by using quantum entanglement  to generate a  communication channel, but there is a notable difference from the standard case:   
  The teleported information appears to be available at System~2
before the measurement of Systems~1 and~3 is actually carried out.
  The interpretation is that the information follows a channel which moves backwards through time in order to be
  teleported outward from  the black hole.

Information flowing backwards in time may sound far fetched.
However,  this protocol is remarkably  similar  to the procedure of post-state teleportation which has been described by Lloyd {\em et al.} 
and shown to avoid all of the usual paradoxes that are associated with time
travel~\cite{lloydPST}. Indeed, post-state teleportation was developed to provide a self-consistent quantum description of time travel, 
inasmuch as general
relativity allows for this possibility through
the existence of closed time-like curves and wormholes.
It was further argued by Lloyd and Preskill
that any issues of  causality and
unitarity violation in the Horowitz--Maldacena protocol would
be small enough to be corrected by considerations
from quantum gravity~\cite{lloydAMPS}.

Lloyd and Preskill also addressed the preservation of monogamy of entanglement.
For instance, the negative-energy particle, System 3, may appear to be entangled
with both System 1 and System 2 --- but, from System 3's perspective, it is only
ever entangled with a single system and  can only ever be sure of which system
is  acting as its purifier when they are in causal contact.

Given that the resolution of the black hole information  paradox
(and firewall problem)   does ultimately depend on wormholes,
one might wonder which of the two discussed solutions
--- ER=EPR and final state --- is the correct one.
After all,  both solutions employ similar procedures of transferring
  information across the black hole horizon via wormholes,
but two distinct solutions to a conundrum is typically one
too many.  To this end, we will employ observer complementarity to shed some light on the situation.
 
\section{Observer Complementarity Revisited}

  Adopting the concept of observer complementarity, we will now argue   
  that ER=EPR and the final-state solution can be viewed as two sides of the same coin rather than two separate solutions
of the black hole information paradox.

  Let us start by considering two observers, Alice and Bob, who are interacting with some black hole. Alice falls into the black hole
  while Bob stays behind to collect
  radiation which is emitted early in the evaporation process. Alice knows nothing about the state of the Hawking radiation that Bob is
  collecting, and so she expects to  experience  flat space while falling through the horizon. The trans-horizon 
entanglement,
which is necessary to ensure a drama-free passage
for Alice,
does not amount to a violation of quantum monogamy because it is the only entanglement that she ever sees. Once inside, Alice
  happens to observe a negative-energy particle (System~3) along with a bit of the original matter (System~1).
Assuming that Alice is (somehow)
  protected  from the tidal forces that are acting deep  within the black hole,
she will see the two in-falling particles heading towards
  annihilation as they approach the singularity. Alice grows concerned that the information held by the in-fallen matter
  will be forever lost inside the black hole. However, using her knowledge of quantum entanglement and post-state selection, Alice soon 
  realizes that, as long as
  the annihilation event acts as a measurement, the information can be teleported backwards in time
  to the positive-energy partner (System~2) which is now moving away from the horizon. Alice then concludes (before
her own violent destruction) that the final-state solution
  allows information to be retrieved from the black hole via a quantum channel of communication and that it does so  without violating
  the condition of monogamy of entanglement.

But, now, what about Bob's perspective?
  Bob has been collecting  early radiation and so is quite aware of the entanglement between  this  and the late radiation
(respectively, $C$ and $B$), which must be
  present to ensure  that the in-fallen information can be retrieved. However, Bob also knows that the equivalence principle should hold at 
  the
  horizon and, as such, the late radiation must be entangled with matter across the horizon (System~$A$). To resolve this apparent conflict,
  Bob concludes that his measurement of the Hawking radiation must have influenced the state of the particles at the horizon ---
   perhaps even producing  a firewall, which would  then thermalize any in-falling matter (including Alice). In order for the influence of 
   these measurements
  to reach the horizon in time, Bob deduces that  a wormhole must connect the interior of the black hole to the Hawking
  radiation.  In this way, Bob comes to the realization that the ER=EPR conjecture is needed to explain the overall procedure.
It can also be noticed that  Bob need not account for Alice's experience
  within the black hole because, as far as he is concerned,  Alice is thermalized  upon entry. Similarly, Alice
  need not account for Bob's actions, which take place outside of her region of causal contact. 

Even though both Alice and Bob interact with the same black hole, their locations on either side of the horizon result
  in much different experiences. However, they can never  compare notes, as Bob cannot reach Alice after she crosses the
  horizon and Alice cannot send a signal that would reach Bob in time
 if he decides to jump in after her (as discussed in $\oint$~2.1).~\footnote{Note
that Alice could not use  post-state teleportation to signal Bob, as this
would be in violation of the so-called unproved-theorem paradox~\cite{lloydPST}.} This
situation may be problematic for our usual notion of  classical physics but is
quite acceptable in the framework of observer complementarity.

One might argue that Alice, as a part of system $A$, plays an essential role in the EPR=ER protocol and thus her perspective 
cannot be discounted when interpreting this proposal as a resolution of the information paradox. Nonetheless, such an argument 
is overlooking the potential of observer complementarity, given that this is indeed a true principle of the fundamental theory 
(for current purposes, we are assuming that it is).  
For  ER=EPR and the final-state solution alike, the role of the interior is to enable information about the initial state
of the black hole to eventually reach
the external radiation (system $C$)
without endangering the entanglement between the pairs (systems $A$ and $B$). Alice and Bob are never in causal contact,
and so the best that either can do is to observe what is happening on their respective side of the horizon and then try to infer
 what is happening on the opposite side. 
Given that the underlying process is quantum teleportation,  the only question left
is if the conduit of the teleported information should be viewed as an Einstein-Rosen bridge or rather as a post-selected measurement. 
Our claim is that it will always be viewed as the former from Bob's perspective and the latter from that of Alice.
If this appears implausible, it is no more or less implausible than Susskind's original scenario: Whereas Alice is happily alive 
(until the tidal forces set in),  Bob is sure that she has already suffered a fiery death. 

The previous results can be summarized as follows: From inside the
  black hole, information is teleported out and preserved via the final-state protocol whereas, outside the black
  hole, information about any disturbance is teleported  inward so that quantum monogamy is preserved via
the ER=EPR mechanism.
  In both cases, there is a quantum 
communication channel that enables the information in question to propagate. Any
difference of opinion
  lies only in the position of the observer, inside or outside the horizon. However, observer complementarity
makes it clear that such differing opinions are par for the course.
 
\section{Conclusion}

  The black hole information paradox and its recent ``firewall'' development can be resolved
by using the notion of wormholes as quantum channels of communication. As reviewed here, two  procedures
that describe just such a resolution
solutions are the ER=EPR argument and the final-state solution. Although
  these are understood as two distinct resolutions, we have argued here that  the ER=EPR and  final-state
solutions can be viewed as precisely the same proposal, only
  from two different perspectives.
The key to our argument is the quantum-gravity inspired  principle of observer complementarity; namely,
that two observers can disagree on events  provided that  they remain out of causal contact.
Ironically, the same basic principle (in the guise of horizon complementarity)
was long thought to provide the answer to the information paradox, until it was recently shot down by the proponents
of the firewall. With apologies to  Mark Twain, the demise of observer complementarity may have  been greatly exaggerated.

\newpage

\section*{Acknowledgments}

The research of KLHB is supported by the NITheP and NRF Bursary Programs.
The research of AJMM is supported by NRF Incentive Funding Grant 85353
and Competitive Programme Grant 93595.
Both authors thank Rhodes University for additional funding and support.

\newpage

\begin{appendices}

\section{The No-Cloning Theorem}
\label{AppA}

  The no-cloning theorem is essentially a restriction on the possibility of producing two 
  identical quantum states from a starting point of one state. 
  The theorem was outlined in 1982 in \cite{nogo} and \cite{nogo2}, and it applies to any 
  general quantum state.
  The proof involves looking at two quantum states which share a Hilbert space. The argument 
  that follows then focuses on the question of what operations could be performed on a system 
  which combines the two states into a tensor product without specifying either state. The 
  application to general 
  states relies on leaving the state that we wish to  copy as an arbitrary unknown state. 
  
   This use of an unknown state is similar to the procedure used in quantum teleportation, as in 
  \cite{plenio}, but there is a fundamental difference to keep in mind. At the end of the 
  teleportation procedure, the unknown state has been transfered to the secondary particle 
  while being  ``destroyed'' at the original particle. So that, when the teleportation procedure is complete, 
  only one particle holds the unknown state. 

  In the no-cloning theorem, however, the question 
  under scrutiny is the possibility of producing two copies of an unknown state which exist 
  simultaneously in two particles. 
  With that aim in mind, possible operations which might result in such a cloned state are 
  considered. The use of a measurement operation is ruled out as it will result in 
  a changed state after the procedure. This leaves the possibility of using an unitary 
  operator on the tensor product which might clone the unknown state. What is seen in \cite{nogo}
  and \cite{nogo2} is that there is no unitary operator which can clone a general unknown state
  from one particle to another in order to end up with two copies of the state in question. 
  
  This conclusion effectively ensures that no operation performed on a system of two particles
  can produce the same state in both particles. In terms of the situation described in the 
  black hole scenario, this restricts what happens to the information being held by the state that
  crosses the horizon. By the no-cloning theorem, this state could  appear inside the 
  horizon or outside, but it cannot be in both places  on the same spacelike slice
  as this would be an example of a cloned 
  state.

\section{The Equivalence Principle}
\label{AppB}

  The original equivalence principle is a reference to an idea which was first derived
  in \cite{ein}. The concept behind the principle is the matching of the effects found 
  in gravitational fields 
  with effects produced in accelerating reference frames. This is most commonly illustrated
  with a comparison of experiments done in a rocket at rest on Earth and similar experiments 
  in the same rocket accelerating through empty space with a force equal to that
  of Earth's gravitational pull.
  This idea was further developed in \cite{ein2} and promoted to the status of a principle of 
  the theory of general relativity. From this idea, Einstein reached the conclusion that the 
  experience of free fall should be indistinguishable from the experience in an inertial
  reference frame for an experimenter inside a closed laboratory. Essentially, if an 
  observer was placed in a closed room, he or she could expect the same results from experiments
  whether that room was placed in free fall around a large mass or if the room was placed in 
  a weightless environment. 
  This concept is often referred to as the ``weak'' equivalence principle. Two further principles 
  have since been developed from it. One, the ``strong'' equivalence principle, relates the above idea to 
  a general range of scenarios and is more encompassing. 
  The second is called Einstein's equivalence principle and it relates specifically to scenarios
  affected by gravity. In essence, it is the same concept as stated above --- that the effects
  felt in an inertial frame are no different from those felt in free fall --- but it clarifies that
  these effects are independent of the free-falling object's location or velocity.

  The relevance for this in the black hole scenario is due to the presence of a large mass producing 
  a substantial gravitational field. An observer falling into the black hole would experience the exact 
  situation that Einstein's equivalence principle applies to. As outlined in \cite{suss}, the 
  curvature of spacetime at the horizon of a massive black hole would be gentle. As such, the free-falling
  observer would experience no tidal forces until he or she was further inside. This means that 
  the experience at the horizon, one of free-fall, should be similar to that experienced 
  in an inertial reference frame which is characterized by empty space.

\section{Quantum Monogamy}
\label{AppC}
  
  {\em The following has been adapted from \cite{thesis} }
  
    This refers to
    the condition that a quantum system may not be strongly entangled with multiple
    systems. This condition is a corollary of the strong-subadditivity statement that was
    proven in \cite{ssa}.     
    Strong subadditivity refers to an inequality that governs how the entropy
    of a system must be constrained with regard to the entropy of the subsystems 
    which make up the whole. By tracing over individual subsystems, the entropies
    of specific sections of the system may be measured and then compared to one another.    
    
    The corollary in question --- that relating  the strong-subadditivity statement to quantum 
    monogamy --- was proven in \cite{araki} and  can be expressed as 
    \begin{equation}
     \label{eq:qm}
     S(\rho^A)+S(\rho^B)\;\leq\;S(\rho^{AC})+S(\rho^{CB})\;.
    \end{equation}
     Here, $S(\rho)$ denotes the entropy of the subsystem described by density matrix $\rho$, and the 
    superscripts $A,B$ and $C$ refer to three subsystems within a larger 
    system. 
    The entropy is compared between the subsystems. This is accomplished by tracing
    out either one or two subsystems from the total density matrix of the complete 
    system. 
    This allows equation \eqref{eq:qm} to 
    limit possible entanglements within a group of three subsystems. 

     Now consider a 
    situation in which subsystem $C$ is strongly entangled with both subsystem $A$ and 
    subsystem $B$. A strongly entangled system has low entropy, and each subsystem 
    within the entangled system will have an individual entropy that is higher than
    the entropy of the entire entangled system. This results in $S(\rho^A)$ being 
    greater than $S(\rho^{AC})$ and, similarly, $S(\rho^B)$ would be greater than 
    $S(\rho^{CB})$. This combination violates equation \eqref{eq:qm} as the left-hand
    side, comprised of single-subsystem entropies, outweighs the right-hand side 
    which consists of the entropies of entangled pairs. This outcome led to the 
    conclusion that system $C$ can only be strongly entangled with either system $A$ 
    {\em or} system $B$ but not both. Hence a quantum system must 
    respect monogamy and may entangle strongly with only one other system at a time.

\section{On Einstein-Rosen Bridges}
\label{AppD}

  {\em The following has been adapted from \cite{thesis}}
  
  The concept that entangled objects can be connected via a wormhole is relevant to 
  the EPR argument in that it allows for actions at Alice's location to disturb Bob's
  system even though the experimenters are separated by spacelike distances. 
  If an entangled pair from  the standard EPR setup is linked by a wormhole, then 
  an action on one of the pair can be felt by the other. Any disturbance caused by 
  Alice's measurement on her system could then  be transmitted through such a wormhole to 
  influence the system at Bob's location. By assuming that entangled pairs are linked
  in this way, the ER=EPR argument provides a mechanism through which the entangled
  pairs maintain their entangled correlations without requiring the spin directions
  to be determined when the particles are prepared. Alice's measurement will still 
  result in a probabilistic outcome consistent with quantum mechanics. The wormhole 
  allows the measurement at Alice's location to influence Bob's system over space-like
  distances; thus providing  a mechanism for Alice's result to influence Bob's result 
  instantaneously regardless of the distance between them. Bob's system would 
  therefore be influenced by the actions at Alice's location. This essentially 
  describes a mechanism which allows for ``spooky action at a distance'' between 
  entangled particles.

\end{appendices}


\begin{thebibliography}{99}






\bibitem{Hawk_info}
  S.~W.~Hawking,
  ``Breakdown of Predictability in Gravitational Collapse,''
  Phys.\ Rev.\ D {\bf 14}, 2460 (1976).


\bibitem{suss} L. Susskind and L. Thorlacius  and J. Uglum, 
``The Stretched Horizon and Black Hole Complementarity,''
Phys. Rev. D {\bfseries 48}, 3743 (1993).




\bibitem{amps}
A.~Almheiri, D.~Marolf, J.~Polchinski and J.~Sully,
      ``Black Holes: Complementarity or Firewalls?,''
       JHEP {\bf 1302}, 062 (2013).



\bibitem{maldacenaSuss}
J.~Maldacena and L.~Susskind,
  ``Cool horizons for entangled black holes,''
  Fortsch.\ Phys.\  {\bf 61}, 781 (2013).


\bibitem{hawk74} S. W. Hawking, ``Black hole explosions,''  Nature {\bf 248}, 30  (1974);
``Particle creation
by black holes,''
Comm. Math. Phys. {\bf 43}, 199 (1975).



\bibitem{horow}
G.~T.~Horowitz and J.~M.~Maldacena,
  ``The Black hole final state,''
  JHEP {\bf 0402}, 008 (2004).





\bibitem{lloydAMPS}
S. Lloyd and J. Preskill,
``Unitarity of black hole evaporation in final-state projection models,''
JHEP {\bf 1408}, 126 (2014).



\bibitem{lloydPST}
S. Lloyd, L.  Maccone,  R. Garcia-Patron, V. Giovannetti and Y. Shikano,
``Quantum mechanics of time travel through post-selected teleportation,''
Phys. Rev. D
{\bf 84}, 025007 (2011).




\bibitem{nogo}
W. Wootters and W.  Zurek,``A Single Quantum Cannot be Cloned," Nature {\bf 299}, 802 (1982). 


\bibitem{bousso}
R.~Bousso,
  ``Positive vacuum energy and the N bound,''
  JHEP {\bf 0011}, 038 (2000).



\bibitem{itzhaki}
N. Itzhaki,
  ``Is the black hole complementarity principle really necessary?,''
  arXiv preprint hep-th/9607028 (1996).



\bibitem{mathur1}
S.~D.~Mathur,
  ``What Exactly is the Information Paradox?,''
  Lect.\ Notes Phys.\  {\bf 769}, 3 (2009).


\bibitem{mathur2} S. D. Mathur,
``The information paradox: a pedagogical introduction,''
Class. Quant. Grav. {\bf 26}, 224001 (2009).




\bibitem{braunstein2009}
S.~L.~Braunstein, S.~Pirandola and K.~Zyczkowski,
  ``Entangled black holes as ciphers of hidden information,''
  Phys. Rev. Lett. {\bf 110}, 101301 (2013).









\bibitem{Unruh}
W. G. Unruh, ``Notes on black-hole evaporation," Phys. Rev. D {\bf 14}, 870 (1976).



\bibitem{araki}
H. Araki and E. H. Lieb,
  ``Entropy inequalities,''
  author={Araki, Huzihiro and Lieb, Elliott H},
  Comm. Math. Phys. {\bf 18}, 160 (1970).
.



\bibitem{EPR}  A. Einstein, B. Podolsky and N. Rosen,  ``Can Quantum-Mechanical Description of Physical Reality Be Considered Complete?,'' 
Phys. Rev. {\bfseries 47}, 777 (1935).





\bibitem{plenio}
M. B. Plenio and V. Vedral,
  ``Teleportation, entanglement and thermodynamics in the quantum world,''
Contemporary physics {\bf 39}, 431 (1998).


\bibitem{PSS}
Y. Aharonov, D. Albert and L. Vaidman,
``How the Result of a Measurement of a Component of a Spin 1/2 Particle Can Turn Out to Be 100?,''
Phys. Rev. Lett. {\bf 60}, 1351 (1988).


\bibitem{ssa}
Lieb, Elliott H., and Mary Beth Ruskai. "Proof of the strong subadditivity of quantum-mechanical entropy." 
In Inequalities, pp. 63-66. Springer Berlin Heidelberg, 2002.

\bibitem{thesis}
K. L. H. Bryan,  ``The EPR Paradox: Back from the Future,'' MSc Thesis, Rhodes University, 2015. 


\bibitem{nogo2}
D. G. B. Dieks, ``Communication by EPR devices,'' Phys. Lett.  A {\bf 92} 271 (1982).

\bibitem{ein}
A. Einstein, ``On the relativity principle and the conclusions drawn from it,''
Jahrbuch der Radioaktivität und Elektronik {\bf 4} 411 (1907).

\bibitem{ein2}
A. Einstein, ``On the Influence of Gravitation on the Propagation of Light,'' 
Annalen der Physik {\bf 35}, 898 906 (1911).

\end{thebibliography}
\end{document}